\documentclass[aps,prl,twocolumn,showpacs,epsfig]{revtex4}
\usepackage{amsmath}
\usepackage{graphicx}
\usepackage{amssymb}

\begin{document}

\title{Itinerant metamagnetism induced by electronic nematic order}
\author{Hae-Young Kee$^{1}$}
\author{Yong Baek Kim$^{1,2}$}
\affiliation{$^{1}$Department of Physics, University of Toronto, Toronto,
Ontario, Canada M5S 1A7 \\
$^{2}$School of Physics, Korea Institute for Advanced Study, Seoul 130-722,
Korea}

\date{\today}

\begin{abstract}

It is shown that metamagnetic transition in metals can occur via 
the formation of electronic nematic order. We consider a simple model 
where the spin-dependent Fermi surface instability gives rise to 
the formation of an electronic nematic phase upon increasing the applied 
Zeeman magnetic field. This leads to two consecutive metamagnetic transitions 
that separate the nematic phase from the low-field and high-field 'isotropic' 
metallic phases. Possible connection to the physics of the bilayer Ruthenate, 
Sr$_3$Ru$_2$O$_7$, is also discussed.

\end{abstract}


\maketitle

{\bf Introduction};
In terms of structural properties of matter,  liquid crystals of molecules have
intermediate forms between the solid and liquid phases. Common examples
of these 'soft' forms of matter include the smectic and nematic phases that 
have spontaneously broken unidirectional translational 
and rotational symmetries, and 
broken rotational symmetry, respectively \cite{deGennes}. 
It is much less appreciated, however, that strongly 
interacting electron systems can give rise to {\it electronic} liquid crystal phases 
where the electronic structure rather than the structural properties of the 
system is 'soft' \cite{kivelson98}. For instance, in two-dimensional electron 
systems, the Wigner crystal and homogeneous electron liquid phases arise in 
the small and high density limits. 
One can ask whether an intermediate form of electronic 
structure arises in this case.
The discovery of quantum Hall nematic phase in two-dimensional electron
gas (2DEG) in high Landau levels \cite{lilly99} clearly shows that the answer is 
affirmative. In this system, the broken rotational symmetry of the nematic
phase is demonstrated in a series of transport measurements \cite{lilly99}.

When electrons reside in the lattice, broken translational and rotational
symmetries would correspond to the broken point group symmetry of the underlying
lattice. It has been suggested that some of these electronic smectic and
nematic phases may play an important role in the physics of strongly correlated 
electron systems \cite{kivelson98}. 
While one may try to understand these phases by following the 
studies of structural liquid crystals, namely introducing topological defects in the 
smectic and nematic phases, it turns out that a different approach may be more
useful in the case of electronic systems. This amounts to studying
the Fermi surface instability of electrons upon tuning the interaction strength
in a novel channel and/or other parameters of the 
system \cite{vadim01,metzner00}. 

An effective model in this spirit that incorporates the two-body interaction 
between quadrupolar density of electrons has been proposed and 
analyzed \cite{vadim01,kim02,kee03_1,kee03_2,kee04}.
It was shown that, in the lattice, the transition to the nematic phase 
as a function of chemical potential is 
generically first-order \cite{kee03_2,kee04}.
This leads to abrupt changes in various physical properties including
the Fermi-surface-topology change from a closed to an open Fermi surface
and Hall constant \cite{kee03_2,kee04}. 
On the other hand, there have been several studies of 
extended Hubbard models that reveal the existence of nematic phases in 
some parameter space of the 
models \cite{metzner00,wegner03,kampf03,metzner03}. 
The direct connection between 
two approaches is not known, but as will be mentioned later the effective 
quadrupolar model seems to be consistent with the conditions under which 
the nematic phases are seen in the studies of extended Hubbard models. 

In this paper, we consider itinerant metamagnetism arising from the formation of 
a nematic phase. Metamagnetism in metals refers to a jump in magnetization from
a low magnetization state to a high magnetization state as an external magnetic field
increases. The transition is first order and associated with a jump in the 
relative volume of the spin-up and spin-down Fermi surfaces. 
In principle, this transition does not have to involve spontaneous breaking of 
the discrete rotational symmetry of the lattice. Here we show that metamagnetic 
transition in metals can occur due to the formation of a nematic phase where only the 
spin-up (or spin-down) Fermi surface spontaneously breaks the discrete 
rotational symmetry.

We study a generalized effective Hamiltonian on the square lattice with the 
quadrupolar-density interaction including the spin degree of freedom.  
As the magnetic field increases, the Fermi surface volume of the 
spin-up Fermi surface increases and when the spin-up Fermi surface gets 
close to the van-Hove singularity, the nematic distortion of the spin-up Fermi 
surface occurs via a first-order transition. 
This leads to the abrupt change in the spin-up Fermi-surface topology from
a closed to an open Fermi surface while the spin-down Fermi surface
changes continuously.  As emphasized in Ref \cite{kee04}, this is an example
where the so-called Lifshitz transition \cite{lifshitz60} of 
non-interacting systems, which
involves a continuous quantum transition between two states with topologically 
distinct Fermi-surface structures without breaking any lattice symmetry,  
is avoided in interacting systems.

When the magnetic field increases further, 
another first-order transition occurs; the spin-up Fermi surface changes from
an open to a closed Fermi surface. The discrete rotational symmetry is now 
restored in the high-field phase. It is shown that two consecutive first-order 
transitions accompany the jumps in the magnetization, leading to 
two metamagnetic transitions.
The transition turns into second order \cite{kee04}  at some finite 
temperature and the nematic phase is bounded above by second 
order transition.
This behavior is reminiscent of that suggested for the unusual
phase near the metamagnetic quantum critical end point of 
Sr$_3$Ru$_2$O$_7$ \cite{perry01,grigera01,grigera04} and
we will discuss possible application of our model to this material. 

{\bf Effective Hamiltonian};
The effective Hamiltonian is based on symmetry considerations
such that the resulting order parameter theory has the correct symmetries. 
In the case of classical liquid crystals, the directions parallel and 
anti-parallel to the rod-like molecule do not make difference.  
As such, the nematic order parameter can be represented as a 
quadrupolar (second order symmetric traceless) tensor built from 
the spatial directions.  
In two-dimensions, it changes sign under a rotation by $\pi/2$ and 
is invariant under a rotation by $\pi$.
In similar spirit, one can construct a quadrupolar order parameter for
electronic systems using the momentum operators of electrons, {\it i.e.},
${\hat Q}_{ij} = {\hat p}_i {\hat p}_j -  {1 \over 2} {\hat p}^2 \delta_{ij} $.
The attractive interaction between quadrupolar densities will lead to
the alignment or anti-alignment of electronic momenta (they are equivalent
just like the case of classical liquid crystals), 
namely the formation of the nematic order.

The simplest generalization of this idea to the electrons with spins on 
the square lattice leads to the following effective Hamiltonian.
We also introduce the external magnetic field, $H_{\rm ext}$, so that 
the effective Hamiltonian only has the Ising degree of freedom in the 
spin space.
\begin{eqnarray}
H &=&  \sum_{{\bf k}, \alpha=\uparrow,\downarrow}
(\epsilon_{\bf k} - \mu) c^{\dagger}_{{\bf k} \alpha} c_{{\bf k} \alpha}
- h \sum_{\bf k} (c^{\dagger}_{{\bf k} \uparrow} c_{{\bf k} \uparrow}
- c^{\dagger}_{{\bf k} \downarrow} c_{{\bf k} \downarrow}) \cr
&+& \sum_{{\bf q}, \alpha=\uparrow,\downarrow} F_2({\bf q})
 \{ {\rm Tr} [ {\hat Q}_{\alpha} ({\bf q})  {\hat Q} _{\alpha}(-{\bf q})] + h.c. \} \ ,
 \end{eqnarray}
where $h = \mu_B H_{\rm ext}$ and $\mu_B$ is the Bohr magneton.
Here $\epsilon_{\bf k}$ is the single particle dispersion and $\mu$ is
the chemical potential. On the square lattice, $\epsilon_{\bf k}
= - 2 t (\cos k_x + \cos k_y)$.  As discussed in Ref. \cite{kee04}, 
the inclusion of 
the next-nearest hopping does not change the qualitative picture, so
we will not include this term for simplicity.
${\hat Q}({\bf q})$ is the lattice analog of the quadrupolar density 
tensor \cite{vadim01}
constructed from the momentum operators and has the following form
in the square lattice.
\begin{eqnarray}
{\hat Q}_{\alpha}({\bf q}) 
= \sum_{\bf k} c^{\dagger}_{{\bf k}+{\bf q} \alpha}
\left ( 
\begin{matrix}
\cos k_x - \cos k_y & \sin k_x \sin k_y \cr
\sin k_x \sin k_y & \cos k_y - \cos k_x \cr
\end{matrix} 
\right ) c_{{\bf k} \alpha} \ .
\end{eqnarray}
Here the interaction, $F_2({\bf q})$, can be any arbitrary short-ranged 
interaction such that
$F_2({\bf q}) \rightarrow -F_2 < 0$ in the $q \rightarrow 0$ limit. 

Some remarks on this model are in order.
In the studies of extended Hubbard models, it was found that the nematic
order (or Pomeranchuk instability) arises when there exist  
attractive (repulsive) interaction between electrons in the same 
(different) patches via forward
scattering \cite{metzner00,kee04}. 
This is consistent with our effective Hamiltonian where the form factor
of the interaction, $-F_2 (\cos k_x - \cos k_y)(\cos k'_x - \cos k'_y)$ provides an
attractive 
interaction between electrons from the same region near $(\pm \pi,0)$, and 
a repulsive interaction between an electron near $(\pm \pi,0)$ and 
another near $(0,\pm \pi)$. Thus it appears that our effective Hamiltonian may
be related to the low energy limit of extended Hubbard models in some 
parameter regime.

{\bf Mean Field Theory};
It is sufficient to concentrate on the $q \rightarrow 0$ limit of the interaction
by letting $F_2({\bf q}) \rightarrow - F_2 \delta_{{\bf q},0}$ with $F_2 > 0$.
In order to obtain mean field theory, the quartic interaction is decoupled
via the following order parameters.
\begin{equation}
\Delta_{\alpha} = F_2 \langle {\hat Q}^{\alpha}_{11} (0) \rangle,  \hskip 0.3cm 
\Delta'_{\alpha}=F_2 \langle {\hat Q}^{\alpha}_{12} (0) \rangle \ .
\end{equation}
The effective single particle dispersion for each spin degree of freedom
now becomes
\begin{equation}
\epsilon'_{{\bf k} \alpha} = -2 t (\cos k_x + \cos k_y) + 
\Delta_{\alpha} (\cos k_x - \cos k_y)
- \Delta'_{\alpha} \sin k_x \sin k_y \ .
\end{equation}
When a non-zero solution for $\Delta_{\alpha}$ and $\Delta'_{\alpha}$ exists,
the spin-$\alpha$ Fermi surface will spontaneously break the discrete rotational 
symmetry of the lattice. It turns out that $\Delta'_{\alpha}$ is always zero in our simple 
model so that the deformation of the Fermi surface occurs along the symmetric
axes of the underlying lattice. Thus we will set $\Delta'_{\alpha}=0$ from now on.

The grand canonical free energy density in the mean field theory,
$F (\mu, h, \Delta_{\uparrow}, \Delta_{\downarrow}) 
= F_{\uparrow} + F_{\downarrow}$, is given by
\begin{equation}
F_{\alpha} (\mu_{\alpha},\Delta_{\alpha})= 
{1 \over F_2}{\Delta^2_{\alpha} \over 2} 
+ F^0_{\alpha} (\mu_{\alpha}, \Delta_{\alpha}) \ ,
\end{equation}
where $\mu_{\uparrow} = \mu + h$ and  $\mu_{\downarrow} = \mu - h$.
Here $F^0_{\alpha}$ is 
\begin{equation}
F^0_{\alpha} (\mu_{\alpha},\Delta_{\alpha})= 
- T \int d\epsilon_{\alpha} \ {\cal D}(\epsilon_{\alpha}) 
\ln (1 + e^{-(\epsilon_{\alpha} - \mu_{\alpha})/T})  \ ,
\end{equation} 
where ${\cal D}(\epsilon_{\alpha})$ is the density of states of the
renormalized single particle dispersion relation for each spin degree
of freedom and has the following functional form.
\begin{equation}
{\cal D} (\epsilon_{\alpha}) = N_0 {\rm Re} \left \{
\frac{1} { \sqrt{1 - \left (\frac{\epsilon_{\alpha}}{4t} \right )^2}} 
K \left ( 1 - \frac{\Delta_{\alpha}^2 - \left ( \frac{\epsilon_{\alpha}} {2} \right )^2} 
{(2t)^2 -  \left ( \frac{\epsilon_{\alpha}}{2} \right )^2} \right ) \right \} ,
\end{equation} 
where $K(x)$ is the elliptic integral of the first kind and $N_0=1/(2\pi^2t)$.
Note the persistence of the logarithmic van Hove singularities 
at $(\epsilon_{\alpha}/2)^2 = \Delta^2_{\alpha}$ that are remnants of
the van Hove singularity of non-interacting systems. 
This leads to non-analytic behavior of the free energy at
$(\mu_{\uparrow}/2)^2= \Delta^2_{\uparrow}$ or
$(\mu_{\downarrow}/2)^2= \Delta^2_{\downarrow}$.

As the external magnetic field increases, $\mu_{\uparrow}$ ($\mu_{\downarrow}$)
increases (decreases) and when the spin-up Fermi surface gets close to 
the first  Brillouin zone boundary, 
the non-analytic behavior of $F_{\uparrow}$ near the van Hove filling,
$(\mu_{\uparrow}/2)^2= \Delta^2_{\uparrow}$, leads to the first order transition to 
the nematic phase or the abrupt deformation of the 
spin-up Fermi surface. Here the spin-down Fermi 
surface is far away from the van Hove filling so that the spin-down Fermi
surface only changes gradually.

{\bf Nematic Order and Magnetization};
At the zero temperature, the free energy, $F = F_{\uparrow} + F_{\downarrow}$
can be further simplified when $\Delta_{\alpha}, \mu_{\alpha}/2 < 2t$ 
and $F_{\alpha}$ for each spin degree of freedom has the following form \cite{kee04}.
\begin{eqnarray}
\hspace{-0.5cm} F_{\alpha} &=& 
\left ( {1 \over F_2} + 2 N_0\right ) {\Delta^2_{\alpha} \over 2} \cr
&+& N_0 \left ( \Delta_{\alpha} + \frac{\mu_{\alpha}}{2}  \right )^2 
\ln \left | \frac{\Delta_{\alpha} + \frac{\mu_{\alpha}}{2}} {4} \right | 
+ (\mu_{\alpha} \rightarrow -\mu_{\alpha})
\end{eqnarray}
where all energy scales are in units of $2t$ for simplicity
and $\Delta_{\alpha}$-independent terms are dropped.

The free energy is investigated as a function of the applied 
magnetic field, $h$, for a given chemical potential $\mu$.
We start  with a closed Fermi surface. As the magnetic field, $h$, 
increases, the spin-up (spin-down) Fermi surface volume increases 
(decreases). For small magnetic fields, there is no spontaneous breaking 
of the lattice symmetry in the Fermi surfaces. 
When $h$ reaches the value that makes the spin-up Fermi surface 
gets close to the van Hove singularity, the nematic order parameter for
the spin-up electrons jumps to a finite value, $\Delta_{\uparrow} \not= 0$.
This represents a first order transition from a closed to an open 
spin-up  Fermi surface. As $h$ further increases, another first order 
transition occurs such that $\Delta_{\uparrow}$ abruptly jumps down 
to zero and the lattice symmetry is restored. 
To summarize, a nematic phase exists in $h_1 < h < h_2$ and 
is bounded by two first order transitions from (to) the low (high) field
'isotropic' phases. 

The behavior of the nematic order parameter for the spin-up Fermi surface 
as a function of $h$ is shown in Fig.1. Here we choose $\mu/(2t)= -0.16$
and $F_2 N_0 = 0.1$. Notice the order parameter jumps to a finite value
at $h_1=0.0428$ and jumps down to zero at $h_2=0.277$. 
\begin{figure}
\includegraphics[height=4.7cm,width=5.7cm,angle=0]{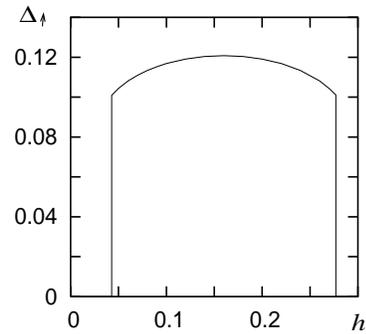} 
\caption{The nematic order parameter, $\Delta_{\uparrow}$, as a function of 
Zeeman magnetic field, $h$, for $F_2 N_0 =0.1$ and $\mu = - 0.16$ in unit
of $2 t$ at $T=0$}
\label{fig:order}
\end{figure}
The behavior of the magnetization, 
$M = \mu_B \sum_{\bf k} (
\langle c^{\dagger}_{{\bf k} \uparrow} c_{{\bf k} \uparrow} \rangle
-  \langle c^{\dagger}_{{\bf k} \downarrow} c_{{\bf k} \downarrow} \rangle )$,
as a function of magnetic field is shown in Fig.2.  
\begin{figure}
\includegraphics[height=7.7cm,width=8.7cm,angle=0]{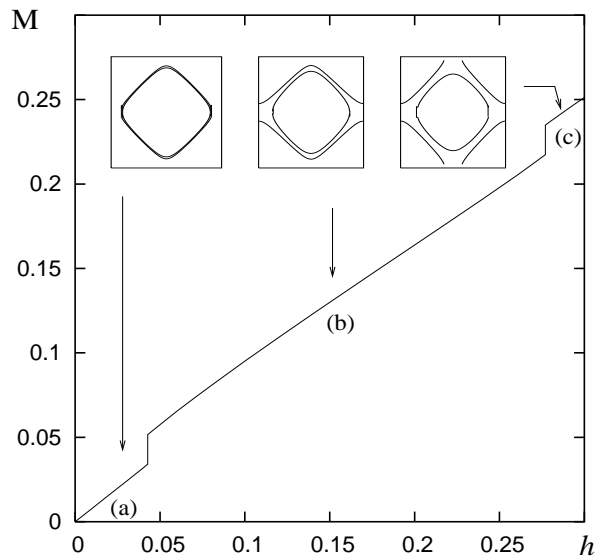} 
\caption{The magnetization, $M$, as a function of Zeeman 
magnetic field for $F_2 N_0=0.1$ and $\mu = - 0.16$. 
Two sharp metamagnetic transitions are due to the phase transitions 
between the 'isotropic', (a) and (c), phases and the nematic state, (b). 
The inset shows the up-spin and down-spin Fermi surfaces
in the nematic and 'isotropic' phases. Note that the down-spin Fermi 
surface volume gradually decreases as the Zeeman magnetic field 
increases, while the up-spin Fermi surface changes abruptly at the 
nematic-'isotropic' transitions.}
\label{fig:metamag}
\end{figure}
Note that the magnetization jumps up at $h_1$ and $h_2$. 
Thus there exist two consecutive metamagnetic transitions.
The corresponding Fermi surfaces in three different regions are shown 
in the inset in Fig.2. In principle, the nematic phase in $h_1 < h < h_2$
will exhibit anisotropic transport properties. In real systems, however,
there will be domains with two possible orientations of the open Fermi
surface for spin-up electrons.

{\bf Possible Applications to Experiments};
Metamagnetism occurs quite often in complex materials, but recent 
experiments on the bilayer ruthenate, Sr$_3$Ru$_2$O$_7$, seem
to provide rather unique opportunity to look at the interplay between
quantum fluctuations and metamagnetism \cite{perry01,grigera01}. 
Early experiments on Sr$_3$Ru$_2$O$_7$
suggest that the metamagnetic critical end-point can be pushed 
down to zero temperature by changing the direction of the 
magnetic field \cite{perry01,grigera01}.
In this case, one expects that the fluctuations near the critical end 
point is inherently quantum mechanical \cite{millis02,kim03}.
Indeed many signatures of quantum critical behavior are seen at finite
temperatures near the putative metamagnetic quantum critical 
end point \cite{perry01,grigera01}.

More recent experiments \cite{perry04}
on much cleaner samples, however, show that
the system avoids the zero temperature quantum critical end point when 
the angle of the magnetic field gets very close to the 'critical' value.
Instead of sustaining large quantum fluctuations, the system undergoes 
two consecutive first order transitions and the magnetization jumps
at each transition \cite{perry04}. 
It has been recently suggested \cite{grigera04} that 
a magnetic field can tune an itinerant system so that the majority-spin
Fermi surface lies close to a van Hove point which may lead to a
spin-dependent Pomeranchuk instability \cite{pomer58} 
which is destroyed by further application of the magnetic field.

At the phenomenological level, our simple model gives rise to
a similar behavior. In the experiment, $h_1$ and $h_2$ are quite
close while our mean field theory gives a relatively wide region
of the nematic phase.  However, the window of the nematic phase
decreases as the value of $F_2 N_0$ decreases.
This quantitative feature can also change if one 
considers a different band structure or introducing an explicit
interaction between spin-up and spin-down electrons. 
On the other hand, we need an attractive
quadrupolar density interaction to get the nematic phase in
a range of magnetic fields while the origin of this interaction 
in real system is not clear. 
Thus it is not easy to make a direct connection between our 
model and the experimental results on Sr$_3$Ru$_2$O$_7$.  
One can imagine, however, that when the metamagnetic fluctuations
become very large, the Fermi surfaces get soft\cite{perry01} and the dominant 
term in the low energy effective Hamiltonian might look like 
what we have studied. 

{\bf Summary and Discussion};
In this paper, we consider a magnetic-field-tuned transition 
at zero temperature between an 'isotropic' (up to the discrete rotational 
symmetry of the lattice) and a nematic phase where the spin-up
(or spin-down) Fermi surface spontaneously breaks the rotational 
symmetry of the lattice.
Such a transition on the lattice is generically first order and 
naturally gives rise to metamagnetism with the jump in the
magnetization across the 'isotropic'-nematic transition.
The first order  transition turns into second order at some finite 
temperature so that the nematic phase is bounded above by 
second order transition\cite{kee04}.

More specifically, we consider a simple model of a quadrupolar
density interaction between electrons in Zeeman
magnetic field. As the magnetic field increases, the Fermi surface 
volume of the spin-up (spin-down) electrons increases (decreases). 
When the spin-up Fermi surface gets close to
the van Hove filling, a first order transition to a nematic phase 
occurs. In the nematic phase, the spin-up Fermi surface spontaneously
breaks the rotational symmetry of the lattice and becomes an
open Fermi surface. As the magnetic field increases further,
another first order transition happens. In the high field phase,
the lattice symmetry is restored and the spin-up Fermi surface 
becomes closed again.
It is shown that the magnetization jumps at each transition
between nematic and isotropic phase.
Note that, in this case, the occurrence of the nematic order is the 
source of the metamagnetism.

In order to obtain these results, we assumed an attractive
interaction between the quadrupolar densities of electrons.
How this interaction can be generated in a more microscopic
model is an interesting question and a subject of future study.

{\bf Acknowledgment}: We are very grateful to Andy Mackenzie, Steve Kivelson,
and Andy Schofield for valuable discussions. 
We would also like to thank  KITP at UCSB and 
Aspen Center for Physics for hospitality, 
where some parts of this work were performed.
This work was supported by the NSERC of Canada, Canadian 
Institute for Advanced Research, Canada Research Chair Program, 
and Alfred P. Sloan Foundation.

\end{document}